\documentclass[letterpaper,10pt]{article} 

\usepackage{opticameet3} 
\newcommand\authormark[1]{\textsuperscript{#1}}
\usepackage{multicol}   
\usepackage{etoolbox}   

\usepackage{amsmath,amssymb}
\newcommand{\ket}[1]{\left|#1\right\rangle}

\usepackage[colorlinks=true,bookmarks=false,citecolor=blue,urlcolor=blue]{hyperref} 

\begin{document}

\title{Emulation of Optically Interconnected Quantum Data Centers Topologies for Cost–Fidelity Benchmarking}


\author{Seyed Navid Elyasi,\authormark{1}, Seyed Morteza Ahmadian,\authormark{1},  Rui Lin,\authormark{1,*} and Paolo Monti\authormark{1}}

\address{\authormark{1}Department of Electrical Engineering, Chalmers University of Technology, Gothenburg, Sweden}

\email{\authormark{*}ruilin@chalmers.se} 


\begin{abstract}
We emulate optically interconnected quantum processors in ring, star, and line topologies using a quantum computer. GHZ benchmarks show that the star provides the best trade-off between cost and fidelity under transduction and fiber noise.
\end{abstract}

\section{Introduction}

\begin{figure*}[b!]
\vspace{-15pt}
\centering
\includegraphics[width=0.8\textwidth]{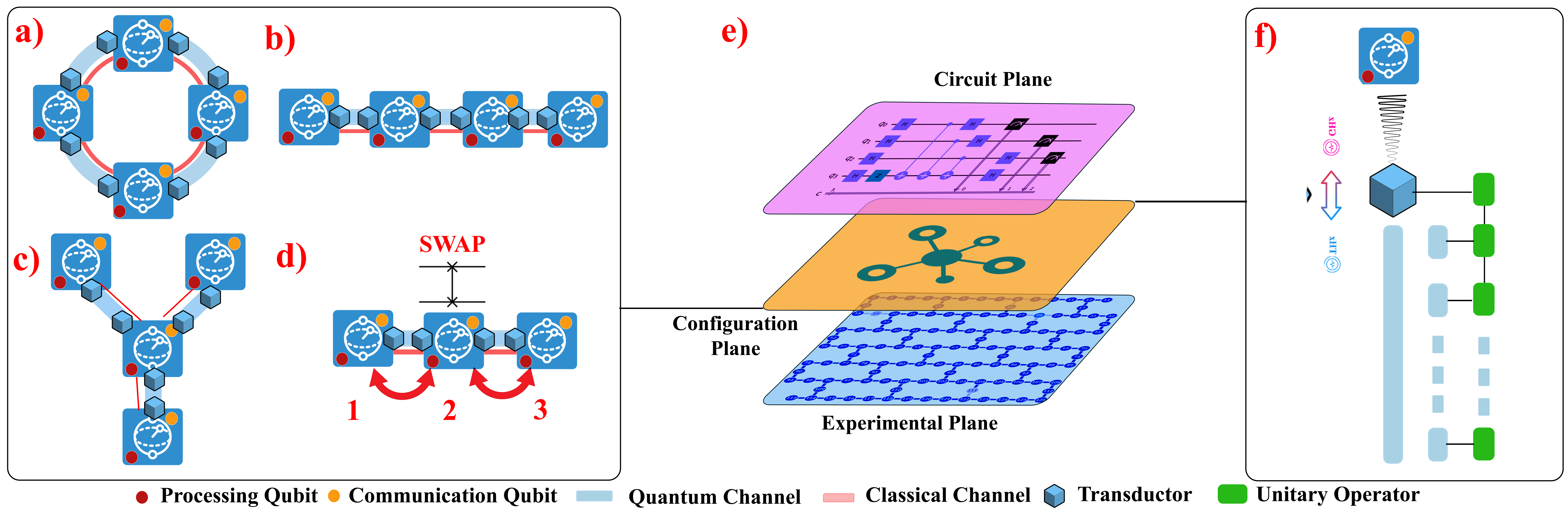}

\caption{(a)–(c) Ring, line, and star QDC topologies. (d) Entanglement distribution via SWAP operations. (e) Simulation layers: circuit, configuration, and experimental layers. (f) Collisional noise model for transduction and fiber effects.}
\label{f:model}
\end{figure*}

Quantum Data Centers (QDCs) address the scalability limitations of current quantum computers, which suffer from cross-talk and limited qubit coherence\cite{campbell2024quantum}. 
In a QDC, multiple Quantum Processing Units (QPUs) are interconnected, typically via optical fibers\cite{aghaee2025scaling}, to enable Distributed Quantum Computing (DQC), quantum sensing, and secure key distribution\cite{liu2024quantum}. 
Achieving quantum advantage in such systems requires generating strong correlations through remote gates(RGs), such as remote CNOTs (RCNOT) between distant qubits. However, interconnect noise arising from transduction and fiber propagation critically limits gate fidelity and, consequently, the overall computational performance.

The interconnection topology of QPUs, as shown in Fig.~\ref{f:model}.(a) to (c), therefore plays a critical role, especially when there is no direct link between two QPUs that need to cooperate, as illustrated in Fig.~\ref{f:model}.(d).  
In such cases, distant entanglement must be established through SWAP operations.  
Here, entanglement is generated between QPUs 1-2, and 2-3. 
To entangle 1 and 3, a SWAP operation must be performed on QPU 2. 
This process introduces additional delay, noise, and qubit decoherence, leading to lower fidelity of multi-hop entanglement. 
Although recent DQC compilers aim to reduce the overhead of remote gates \cite{cuomo2023optimized}, large-scale algorithms, such as Shor’s, still require multi-hop entanglement. This reflects a fundamental trade-off between connectivity and cost: sparse topologies with fewer inter-QPU links reduce hardware complexity but incur higher latency and fidelity degradation due to multi-hop SWAP operations, whereas denser topologies with direct point-to-point links lower communication overhead and noise at the expense of increased interconnect cost.

In this study, we propose an emulation-based approach to explore this trade-off systematically. 
Using a quantum computer, we emulate QDCs implemented in ring, line, and star topologies under communication noise and qubit decoherence. Both the communication cost (i.e., the number of entanglement links) and circuit fidelity used for the generation of GHZ states, which interconnect all the processing qubits in all QPUs, are evaluated.

As illustrated in Fig.~\ref{f:model}.(e), the proposed framework spans three layers: the circuit layer, where the GHZ-state generation circuit is defined; the configuration layer, which distributes the circuit across different topologies while introducing noise from optical and transduction and fiber through a collisional model as shown in Fig.~\ref{f:model}.(f); and the experimental layer, which maps the resulting configuration onto the qubit-coupling layout of a superconducting QPU.

Our results show that a star topology consistently achieves the best trade-off between cost and fidelity, benefiting from direct inter-QPU links that minimize SWAP operations and noise accumulation, while the line topology suffers the most from accumulated decoherence.
Tomographic validation of the remote-CNOT operation confirms the accuracy of the noise model, showing strong transduction-induced degradation but minor optical-fiber contribution to fidelity loss.
Overall, this work provides a practical framework to benchmark and compare QDC interconnection topologies, offering design insights for scalable and optically interconnected quantum computing systems.

\section{System Architecture and Noise Modeling}
\begin{figure*}[t!]
\begin{center}
\includegraphics[width=1\textwidth]{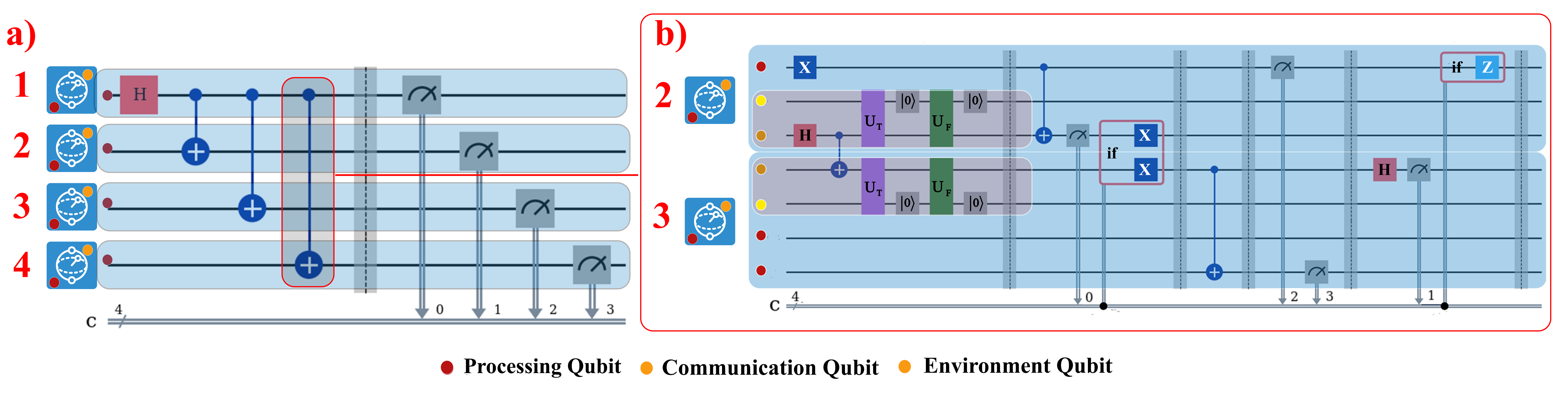}
\end{center}
\vspace{-20pt} 
\caption{(a) GHZ-state circuit connecting processing qubits for QDC evaluation. (b) Cat-comm protocol implementing the remote CNOT gate with noisy entanglement generation.
}
\vspace{-20pt} 
\label{f:circuit}
\end{figure*} 

Each QPU in the QDC architecture (Fig.~\ref{f:model}) comprises processing qubits (red circles) for circuit execution, communication qubits (orange circles) for entanglement generation, and flying qubits—photons used to entangle distant communication qubits. 
As shown in Fig.~\ref{f:model}.(a), the first processing qubit is initialized in superposition via a Hadamard gate, followed by remote CNOT gates applied pairwise between QPUs.

To implement the RCNOT gate, we employ the cat-comm protocol proposed in \cite{campbell2024quantum}. 
Environmental noise is emulated by introducing an additional environment qubit (colored in yellow in Fig.~\ref{f:circuit}(b)) coupled to the communication qubits. 
To capture both transduction and fiber propagation effects, the environment qubit interacts through two types of iterative unitary operations:
$\hat{U}_{T}$ for transduction and $\hat{U}_{F}$ for the optical fiber channel. 
The transduction layer mainly contributes to amplitude damping due to frequency conversion inefficiencies, 
while the optical fiber segment induces weaker effects associated with propagation loss and photon scattering. 
After each iteration, the environment qubit is reset, reducing resource overhead while enforcing a Markovian evolution consistent with the theoretical framework in \cite{elyasi2025framework}.

The operators $\hat{U}_{T}$ and $\hat{U}_{F}$ represent two-qubit time-evolution unitaries of the form 
$e^{-i\hat{H}\Delta t}$, generated by the interaction Hamiltonian 
$\hat{H} = \kappa \left(\hat{\sigma}_{+}\hat{\sigma}_{-} + \hat{\sigma}_{-}\hat{\sigma}_{+}\right)$, 
where $\kappa$ denotes the normalized coupling strength between the environment and communication qubits. 
A higher $\kappa$ is assigned to the transduction interface, which exhibits stronger coupling and higher decoherence. 
At the same time, a smaller value is used for the optical fiber segment, whose weaker interactions mainly contribute to gradual propagation loss.

The proposed circuits are executed on digital quantum computers and programmed using the \textit{Qiskit} Python package. 
All source codes, data, job-ids, and JSON files of the experiments are available in the GitHub repository~\cite{nelyasi_QDC-OFC}. 
Together with the developed noise model, these circuits are emulated across different QDC topologies (i.e., star, ring, and line), each exhibiting a distinct trade-off between the communication cost (in terms of entanglement links) and the achieved fidelity, forming a comprehensive evaluation framework for optically interconnected quantum data centers.

\section{Performance Evaluation of QDC Topologies}

The emulations were executed using the IBM Quantum Torino backend based on the 133-qubit Heron r1 processor, which supports feedforward operations and mid-circuit measurements. 
To map each topology onto the physical qubit coupling graph, we selected subsets of qubits forming the desired connectivity and transpiled the circuits using an initial layout corresponding to these indices (topology configurations are available at \cite{nelyasi_QDC-OFC}).

\begin{figure*}[t!]
\begin{center}
\includegraphics[width=1\textwidth]{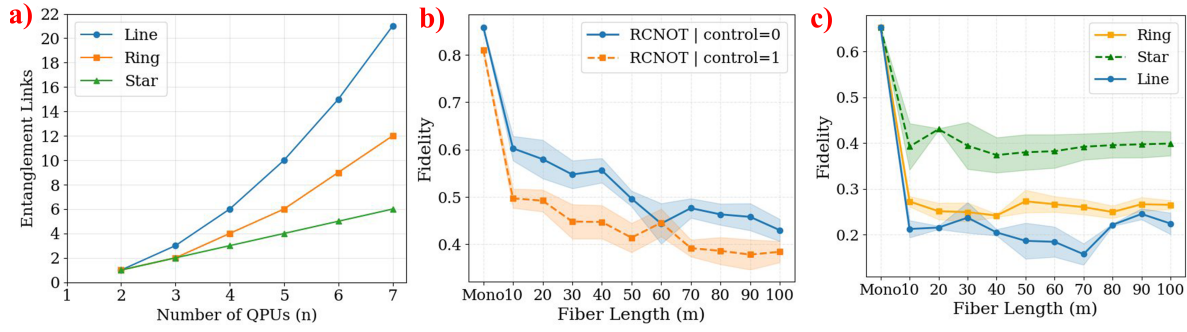}
\end{center}
\vspace{-15pt} 
\caption{
(a) Communication cost for GHZ-state execution across QDC topologies. (b) RCNOT fidelity results based on the noise model. (c) Fidelity results for the GHZ circuit.}
\vspace{-20pt} 
\label{f:results}
\end{figure*}

The communication cost analysis results are shown in Fig.~\ref{f:results}(a). The figure illustrates the number of entanglement links required between QPU pairs for GHZ state generation. The total number of entanglement links is given by $E_{\text{line}} = \tfrac{n(n-1)}{2}$ for a line topology, $E_{\text{ring}} = \tfrac{n^2}{4}$ for even $n$ or $\tfrac{(n-1)(n+1)}{4}$ for odd $n$ for a ring topology, and $E_{\text{star}} = n - 1$ for a star topology. For a four-QPU system, the total interconnect cost follows the order line $>$ ring $>$ star.

We first benchmarked the RCNOT operation under the previously proposed noise model \cite{nelyasi_QDC-OFC} between the processing qubits of a pair of QPUs. 
As shown in Fig.~\ref{f:results}.(b) the monolithic RCNOT gate achieves beyond $80\%$ fidelity for both control states $\ket{0}$ and $\ket{1}$. 
In the distributed implementation, however, the first step includes $10\,\mathrm{m}$ of a G.654.E optical fiber with an attenuation constant of $0.0392~\text{km}^{-1}$. The transduction process—with a coupling constant of $0.5$—has a significant impact on fidelity loss, leading to around $30\%$ degradation, which aligns with reported values for state-of-the-art transducers. 
The optical fiber contribution, in contrast, has a smaller effect, with fluctuations around $50\%$ to $40\%$ fidelity after four hops.

When scaling to multi-QPU GHZ-state generation, distinct behaviors emerge across topologies. 
As shown in Fig.~\ref{f:results}(c), the line topology exhibits the strongest fidelity loss. 
This performance degradation stems from its asymmetric structure and the higher number of required entanglement links (Fig.~\ref{f:results}(a)) for entanglement distribution, which increases circuit depth and causes other qubits to decohere while entanglement is being generated. 
The additional SWAP operations required within the QPUs further amplify this effect. 
In contrast, the star topology consistently achieves the highest fidelity, benefiting from its direct fiber-based connections between the central and peripheral QPUs that minimize both routing and entanglement overhead. 
The ring topology demonstrates intermediate performance between the line and star configurations, requiring fewer entanglement links than the line while providing greater overall connectivity than the star, as illustrated in Fig.~\ref{f:results}(a). 

Overall, our results indicate that the star topology, as expected from its higher pairwise connectivity, 
introduces less noise and achieves the highest overall fidelity for the GHZ circuit. 
Furthermore, the results highlight the flexibility of the emulation framework developed in~\cite{elyasi2025framework}, 
which enables the exploration of various QDC topologies and provides insight into how optical-channel noise and transduction inefficiencies influence the global performance and scalability 
of optically interconnected QDCs.

\section{Conclusions}

We emulated optically interconnected QDCs using GHZ-state generation to compare star, ring, and line topologies. Results show that the star topology offers the best trade-off between fidelity and communication cost. The proposed framework enables benchmarking and design exploration of scalable, optically interconnected quantum data centers.

\section*{\footnotesize Acknowledgments:}\footnotesize 

This work is supported by the project “ DIGITAL-2022-QCI-02-DEPLOY-NATIONAL” (Project number: 101113375 — NQCIS) funded by the EU together with VINNOVA and WACQT. Additional support is provided by VR and GENIE. We thank Chalmers Next Labs for providing a premium account and support for the IBM Quantum platforms.

\bibliographystyle{opticaconf}
\bibliography{sample}

\end{document}